\documentclass[aps,groupedaddress]{revtex4}
\usepackage{epsfig}
\usepackage{graphicx}
\usepackage[T1]{fontenc}
\usepackage{ae}
\usepackage{color}
\usepackage[latin1]{inputenc}
\usepackage{amssymb,amsbsy,amsmath}
\usepackage{bbm}
\usepackage{undertilde}

\newtheorem{lemma}{Lemma}

\newtheorem{prop}{Proposition}

\newtheorem{theorem}{Theorem}
\newtheorem{cor}{Corollary}

\newcommand{\op}[1]{\fontdimen12\textfont3=2pt\fontdimen12\scriptfont3=1.4pt\!\null\mathop{\protect\vphantom{#1}\smash{#1}}\limits_{\sim}\null\!}

\begin{document}



\title[Conditional action]
{Conditional Action and imperfect Erasure of Qubits}

\author{Heinz-J\"urgen Schmidt
}
\affiliation{ Universit\"at Osnabr\"uck,
Fachbereich Physik,
 D - 49069 Osnabr\"uck, Germany
}


\begin{abstract}
We consider state changes in quantum theory due to ``conditional action" and relate these to the discussion
of entropy decrease due to interventions of ``intelligent beings" and the principles of Szilard and Landauer/Bennett.
The mathematical theory of conditional actions is a special case of the theory of ``instruments''
which describes changes of state due to general measurements and will therefore be briefly outlined in the present paper.
As a detailed example we consider the imperfect erasure of a qubit that can also be viewed as a conditional action
and will be realized by the coupling of a spin to another small spin system in its ground state.
\end{abstract}

\maketitle

\section{Introduction}\label{sec:I}

According to a widespread opinion, there are two types of state change in quantum mechanics:
Time evolution in closed systems, and state changes due to measurements. The mathematical description of these two processes is known in principle:
\begin{enumerate}
  \item[(i)] Time evolution in closed systems can be described by means of unitary operators $U(t)$ according to
  \begin{equation}\label{zeit1}
   \rho\mapsto U(t)\,\rho\, U(t)^\ast  \;,
  \end{equation}
  where $U(t)$ is obtained by solutions of the time-dependent Schr\"odinger equation and $\rho$ denotes any statistical operator.
  \item[(ii)] Conditional state changes according to the outcome of a measurement of an observable $A$ will be described, in the simplest case,
  by maps of the form
  \begin{equation}\label{mess1}
   \rho\mapsto P_n\, \rho\,P_n
   \;,
  \end{equation}
  where $\left( P_n\right)_{n\in{\mathcal N}}$ is the family of eigenprojections of a self-adjoint operator
  $A=\sum_n a_n P_n$. Without selection according to the outcomes of the measurement the total state change will be
  \begin{equation}\label{mess2}
   \rho\mapsto   \sum_{n\in{\mathcal N}} P_n\, \rho\,P_n
   \;,
  \end{equation}
\end{enumerate}
In a recent article \cite{S20} we have suggested a third type of state change, called ``conditional action", that
combines the two afore-mentioned ones insofar as it describes a state change depending on the result of a preceding measurement.
\begin{enumerate}
  \item [(iii)] In the simplest case a conditional action is mathematically described by maps of the form
  \begin{equation}\label{condact1}
  \rho\mapsto U_n\,P_n\, \rho\,P_n\,U_n^\ast
  \;,
  \end{equation}
  with the same notation as in (\ref{mess1}) and a family of unitary operators $\left(U_n\right)_{n\in{\mathcal N}}$.
  Without selection according to the outcomes of the measurement the total state change will be
  \begin{equation}\label{condact2}
   \rho\mapsto   \sum_{n\in{\mathcal N}} U_n\,P_n\, \rho\,P_n\,U_n^\ast
   \;,
  \end{equation}
\end{enumerate}

Before explaining the details of this suggestion and suitable generalizations
we will fix some general notation used in the present paper.
A measurement leading to the state change (\ref{mess1}) is called a ``L\"uders measurement"
in accordance with \cite{BLM96}, sometimes also called ``projective measurements" in the literature.
In order not to have to go into technical intricacies the quantum system $\Sigma$ will be
described by a finite-dimensional Hilbert space ${\mathcal H}$.
In this case the index set ${\mathcal N}$ will also be finite.
Let $B({\mathcal H})$ denote the real linear space of Hermitean operators $A:{\mathcal H}\rightarrow{\mathcal H}$,
and $B_1^+({\mathcal H})$ the convex subset of statistical operators, i.~e.,
Hermitean operators $\rho$ with non-negative eigenvalues and $\mbox{Tr} \rho=1$.
The state changes considered above in (\ref{mess1}) and (\ref{mess2}) can be viewed as a map
$\mathfrak{L}:{\mathcal N}\times B({\mathcal H}) \rightarrow B({\mathcal H})$ defined by
\begin{equation}\label{LI1}
  \mathfrak{L}(n)(\rho):=  P_n\, \rho\,P_n
  \;,
\end{equation}
that will be called a ``L\"uders instrument" and the corresponding map
\begin{equation}\label{LI2}
  \mathfrak{L}({\mathcal N})(\rho):=  \sum_{n\in{\mathcal N}} P_n\, \rho\,P_n
\;,
\end{equation}
the ``total L\"uders operation".
A difference of the two state changes according to (i) and (ii) arises when we consider the change of the von Neumann entropy
\begin{equation}\label{vNE}
  S(\rho) := -\mbox{Tr}\left( \rho\,\log \rho\right),\quad \mbox{for  }\rho\in B_1^+({\mathcal H})
  \;.
\end{equation}
Under unitary time evolutions (i) the entropy remains constant,
\begin{equation}\label{Econ}
  S(\rho)=S\left(U_t\,\rho\,U_t^\ast\right)
  \;,
\end{equation}
whereas for L\"uders measurements (ii) the entropy may increase and we can only state that
\begin{equation}\label{Einc}
 S(\rho) \le S(\mathfrak{L}({\mathcal N})(\rho))
 \;,
\end{equation}
see \cite{vN32} -- \cite{SG20}, in accordance with the second law of thermodynamics.

In contrast to closed systems, time evolution in open systems can take a more general form.
An obvious model to account for the time evolution in open systems is to consider the extension of the system $\Sigma$
with Hilbert space ${\mathcal H}$ by another, auxiliary system $E$ (environment, heat bath, measurement apparatus, \ldots)
with Hilbert space ${\mathcal K}$ and the unitary
time evolution $V$ of the total system $\Sigma+E$. If the total system is initially in the state $\rho\otimes \sigma$
it will generally evolve into an entangled state $V\left( \rho\otimes \sigma\right) V^\ast$.
In the end, we again consider the system $\Sigma$ and find its reduced state $\rho_1$ given by the partial trace
\begin{equation}\label{open1}
 \rho_1=\mbox{Tr}_{\mathcal K}\left( V\left( \rho\otimes \sigma \right) V^\ast\right)
 \;.
\end{equation}
The corresponding state change $\rho\mapsto \rho_1$ will, in general, not be of unitary type (\ref{zeit1}),
but represents a natural extension (ie) of the state changes according to (i). In general, the entropy balance
for these state changes is ambivalent:
$S(\rho_1)$ can be smaller or larger than $S(\rho)$. In fact, the initial entropy of the total system
is $S(\rho)+S(\sigma)$ and the unitary time evolution $V$ leaves this invariant. But the separation of the
total system into its parts $\rho_1$ according to (\ref{open1}) and
\begin{equation}\label{open2}
 \rho_2=\mbox{Tr}_{\mathcal H}\left( V\left( \rho\otimes \sigma \right) V^\ast\right)
 \;,
\end{equation}
{\em increases} the entropy (or leaves it constant) according to ``subadditivity" of $S$, see \cite{NC00}, 11.3.4.,
and hence
\begin{equation}\label{open3}
  S(\rho)+S(\sigma)\le S(\rho_1) + S(\rho_2)
  \;.
\end{equation}
But $S(\rho_1)-S(\rho)$ may assume positive or negative values. This can be physically understood as the phenomenon that,
apart from a possible increase of the total entropy according to (\ref{open3}), there may be an entropy flow
from the system $\Sigma$ into the environment $E$ or vice versa.

An analogous extension of the system $\Sigma$ to $\Sigma+E$ can also be considered for L\"uders measurements.
We again start with an initial total state $\rho\otimes\sigma$, where $\sigma\in B_1^+({\mathcal K})$,
and a unitary time evolution $V$ of the total system.
Then a L\"uders measurement corresponding to a complete family $\left( Q_n\right)_{n\in{\mathcal N}}$
of mutually orthogonal projections of the auxiliary system is performed and the post-measurement total state is reduced to the system
$\Sigma$ leading to the final state
\begin{equation}\label{inst1}
 \mathfrak{I}(n)(\rho):=
 \mbox{Tr}_{\mathcal K}\left(\left( \mathbbm{1}\otimes Q_n\right) V\left( \rho\otimes \sigma \right) V^\ast\left( \mathbbm{1}\otimes Q_n\right)\right)
 \;,
\end{equation}
or, without selection, to
\begin{equation}\label{inst2}
 \mathfrak{I}({\mathcal N})(\rho):=
 \mbox{Tr}_{\mathcal K}\left(\sum_n \left( \mathbbm{1}\otimes Q_n\right) V\left( \rho\otimes \sigma \right) V^\ast\left( \mathbbm{1}\otimes Q_n\right)\right)
 \;.
\end{equation}
Thus we obtain extensions (iie) of the state changes (ii) due to L\"uders measurements by maps
$ \mathfrak{I}:{\mathcal N}\times B({\mathcal H}) \rightarrow B({\mathcal H})$ of the form (\ref{inst1}),
that are called ``instruments" in the literature, see \cite{BLPY16} and Section \ref{sec:DR} for more precise mathematical definitions.
L\"uders instruments are idealized special cases of general instruments that, in some sense,
minimize the perturbation of the $\Sigma$ system by the measurement, but ``real measurements"
are better described by general instruments. Analogous remarks as in the case of open systems apply
for the entropy balance:
It is well-known, see \cite{L73} or \cite{NC00}, Exercise 11.15, that general measurements may decrease the system's entropy.

The latter observation has lead us to the suggestion \cite{S20} that the entropy decrease of systems due to the
``intervention of intelligent beings" as, e.~g., Maxwell's demon, can be explained by the same mechanism.
Originally, the notion of ``conditional action" was developed to describe the intervention of Maxwell's demon
in the energy distribution of a gas with two chambers:
Depending on the result of an energy measurement on a gas molecule approaching the partition between the two chambers,
a door is opened or shut. Thus, the further time evolution of the gas depends on the result of the measurement.
Similarly, the result of measuring whether a single molecule is in the left or right chamber
can be used to trigger an isothermal expansion to the right or left (Szilard's engine).
Szilard argues \cite{S29} that the entropy decrease of the system is compensated by the entropy costs of acquiring information about the position
of the gas particle (``Szilard's principle"). His arguments are formulated within classical physics and not easy to understand, see
also the analysis and reconstruction of Szilard's reasoning in \cite{LR94}, \cite{EN98} and \cite{EN99}.
Nevertheless, it seems possible that the entropy decrease due to such external interventions is a special
case of the well-understood entropy decrease due to state changes described by general instruments.

In fact, it can be easily confirmed, that the maps of the form (\ref{condact1}) describing ``conditional action"
are special cases of instruments and hence are called ``Maxwell instruments" in \cite{S20}.
The mathematical notion of state changes described by instruments is sufficiently
general to cover not only changes due to inevitable measurement disturbances but also ``deliberate" state changes
depending on the result of a measurement.

This notion of ``conditional action" will be slightly generalized in the
present paper, and then comprises not only interventions of Maxwell's demon or cycles of Szilard's engine
\cite{S29}, \cite{Z84}, \cite{S20}, but also quantum teleportation \cite{NC00} Ch. 1.3.7, quantum error correction \cite{NC00} Ch. 10
or erasure of qubits \cite{S20}.

Relative to the choice of a suitable basis a qubit has two values, ``0" or ``1". Consider a ``yes-no"-measurement
corresponding to said basis. If the result is ``1" the two states are swapped, hence $``1" \mapsto ``0"$.
If the result is ``0"  then nothing is done,  hence $``0" \mapsto ``0"$. This constitutes the conditional action
which sets the qubit state to its default value ``0" at any case and hence can be legitimately considered as
an ``erasure of a qubit".

The latter example contains an ironic punch line in that the erasure of memory contents
with measurement results and the corresponding entropy costs are usually considered to resolve the apparent contradiction
between the actions of Maxwell's demon and the second law (Landauer's principle).
If memory erasure itself were taken as a conditional action,
we would seem to enter an infinite circle of creating and erasing new memory contents.
The obvious resolution to this problem is the observation that the entropy decrease in the system
$\Sigma$ is due to some flow of entropy from $\Sigma$ to the auxiliary system $E$ as described above.
If $E$ can be viewed as a ``memory device" then, at the end of the conditional action,
it already contains the missing entropy. It is not necessary to erase the content of the memory.
The latter would not create the missing entropy, but only make it visible.

As in \cite{S20} it seems sensible to distinguish between the principle that erasure of memory produces entropy \cite{L61}
(``Landauer's principle" in the narrow sense) and the position that this effect constitutes the solution of the apparent
paradox of Maxwell's demon \cite{B82} (henceforward called ``Landauer/Bennett principle").
Moreover, it will be a matter of substantiating our critique of the Landauer/Bennett principle
(not of the Landauer principle) outlined above with a more realistic model of qubit erasure than that given in \cite{S20}.

To this end we realize the qubit (the system $\Sigma$) by a single spin with spin quantum number $s=1/2$ described by a Hilbert space
${\mathcal H}\cong {\mathbbm C}^{2s+1}={\mathbbm C}^{2}$ and model the erasure of the qubit by the coupling
of the single spin with a ``heat bath" $E$ consisting of $N=6$ spins
such that the time evolution of the total system can be analytically calculated.
The quotation marks refer to the fact that the ``heat bath" is pretty small and not macroscopic,
as usually required, and that, moreover, it is rather a ``cold bath".
This is due to the choice of the default value "0" of the qubit as the ground state $\downarrow$ of the single spin.
Thus, erasing the qubit is physically equivalent to cooling the system $\Sigma$ to the temperature $T=0$.
Although this is, strictly speaking, impossible due to the third law of thermodynamics, it can be approximately
accomplished by coupling the single spin to a system of $N=6$ spins in its ground state. Here we ignore the
physical impossibility to prepare a system in its ground state and consider the ground state of the ``heat bath" as a suitable approximation
to a state of very low temperature.
This approximation has the advantage of providing fairly simple expressions for the relevant quantities considered in this paper.
The corresponding calculations are presented in Section \ref{sec:ER}.

As a side effect of this account results the necessity to define the concept of ``conditional action"
somewhat more generally than in \cite{S20}. This is done in Section \ref{sec:DR} where we also recapitulate the basic
notions of quantum measurement theory required for the present work. A critical account of the Szilard principle
in the realm of quantum theory is given in Section \ref{sec:SP}, where we also formulate an upper bound for the entropy decrease
due to conditional action that is compatible with Szilard's reasoning but only valid under certain restrictions.
A similar bound is derived in Section \ref{sec:OLR} where the connections of the present theory with the OLR approach \cite{A17} are considered.
The proofs are moved to the appendix, as is the explicit construction of a ``standard" measurement dilation for a general instrument.
This measurement dilation is well-known but nevertheless reproduced here since some arguments given in this paper depend on its details.
We close with a Summary and Outlook in Section \ref{sec:SU}.

\section{General definitions and results}\label{sec:DR}

In the following we will heavily rely upon the mathematical notions of {\em operations} and {\em instruments}.
Although these notions are well-known, see, e.~g., \cite{K83} -- \cite{P13b} and  \cite{BLPY16},
it will be in order to recall the pertinent definitions adapted to the present purposes
and their interpretations in the context of measurement theory.
For readability, we sometimes will repeat definitions already presented in the introduction \ref{sec:I}.

Let ${\mathcal H}$ be a  $d$-dimensional Hilbert space,
$B({\mathcal H})$ denote the space of Hermitean operators  $A:{\mathcal H}\longrightarrow {\mathcal H}$
and $B^+({\mathcal H})$ the cone of positively semi-definite operators, i.~e., having only non-negatives eigenvalues.
The convex subset $B_1^+({\mathcal H})\subset B^+({\mathcal H})$ consists of {\em statistical operators}  $\rho$
with $\mbox{Tr} \rho=1$. Such operators physically describe (mixed) states.  {\em Pure states} are
represented by one-dimensional projectors $P_\psi$, where $\psi\in{\mathcal H}$ with $\|\psi\|=1$.

According to \cite{NC00}, 8.2.1,  there are three equivalent ways to define {\em operations}:
\begin{itemize}
  \item By considering the system $\Sigma$ coupled to environment $E$,
  \item by an operator-sum representation, or
  \item via physically motivated axioms.
\end{itemize}
Here we follow the second approach and define an ``operation" to be a map
$A:B({\mathcal H})\longrightarrow B({\mathcal H})$ of the form
\begin{equation}\label{OI2}
 A( \rho)= \sum_{i\in{\mathcal I}}A_i\,\rho\,A_i^\ast
  \;,
\end{equation}
with the {\em Kraus operators} $A_i:{\mathcal H}\rightarrow{\mathcal H}$ and a finite
index set ${\mathcal I}$, see \cite{K83}. It follows that an operation is linear
and maps $B^+({\mathcal H})$ into itself. It may be trace-preserving or not.

Operations are intended to describe state changes due to measurements. For example, the total L\"uders operation
(\ref{LI2}) is a trace-preserving operation in the above sense with
${\mathcal I}={\mathcal N}$ and $A_n=P_n$ for all $n\in{\mathcal N}$.
An operation $A:B({\mathcal H})\rightarrow B({\mathcal H})$ will be called {\em pure} iff
the representation  (\ref{OI2}) of $A$  can be reduced to a single Kraus operator, i.~e.,
\begin{equation}\label{PO2}
  A(\rho) = A_1\,\rho\,A_1^\ast
  \;.
\end{equation}
Physically, this means that a pure operation maps pure states onto pure states, up to a positive factor.

There exists a so-called statistical duality between states and observables, see \cite{BLPY16}, chapter 23.1.
In the finite-dimensional case $B({\mathcal H})$ can be identified with its dual space $B({\mathcal H})^\ast$
by means of the Euclidean scalar product $\mbox{Tr}\,(A\,B)$. Physically, we may distinguish between the two
spaces in the sense that $B({\mathcal H})$ is spanned by the subset of statistical operators representing states
and $B({\mathcal H})^\ast$  is spanned by the subset of operators with eigenvalues in the interval $[0,1]$ representing
{\em effects}. Effects describe yes-no-measurements including the subset of projectors, which are the extremal points of the
convex set of effects, see \cite{BLPY16}.

Every operation $A:B({\mathcal H})\rightarrow B({\mathcal H})$, viewed as a transformation of states (Schr\"odinger picture)
gives rise to the dual operation $A^\ast:B({\mathcal H})^\ast\longrightarrow B({\mathcal H})^\ast$
viewed as a transformation of effects (Heisenberg picture).
Reconsider the representation  (\ref{OI2}) of the operation $A$ by means of the Kraus operators $A_i$. Then the dual
operation $A^\ast$ has the corresponding representation
\begin{equation}\label{M1}
 A^\ast(X)= \sum_{i\in{\mathcal I}}A_i^\ast\,X\,A_i
  \;,
\end{equation}
for all $X\in B({\mathcal H})^\ast$.

Let ${\mathcal N}$ be a finite set of outcomes. Then the map
${\mathfrak I}:{\mathcal N}\times B({\mathcal H})\longrightarrow B({\mathcal H})$
will be called an {\em instrument} iff
\begin{itemize}
  \item ${\mathfrak I}(n)$ is an operation for all $n\in{\mathcal N}$, and
  \item $\mbox{Tr}\left(\sum_{n\in{\mathcal N}}{\mathfrak I}(n)(\rho)\right)=\mbox{Tr}\rho$ for all $\rho\in B({\mathcal H})$.
\end{itemize}
The first condition can be re-written as
\begin{equation}\label{OI3}
  {\mathfrak I}(n)(\rho) = \sum_{i\in{\mathcal I}_n}A_{ni}\,\rho\, A_{ni}^\ast \quad\mbox{for all } n\in{\mathcal N},
\end{equation}
with suitable Kraus operators $A_{ni}:{\mathcal H} \rightarrow {\mathcal H}$.
The second condition can be rephrased by saying that the {\em total operation} ${\mathfrak I}({\mathcal N})$ defined by
\begin{equation}\label{OI4}
  {\mathfrak I}({\mathcal N})(\rho)\equiv \sum_{n\in{\mathcal N}}{\mathfrak I}(n)(\rho)
\end{equation}
will be trace-preserving. An instrument ${\mathfrak I}$ will be called ``pure" iff each operation ${\mathfrak I}(n),\; n\in{\mathcal N},$
is pure.

Examples of pure instruments are given by L\"uders instruments (\ref{LI1}) and ``Maxwell instruments" (\ref{condact1}).

Similarly as for operations, every instrument ${\mathfrak I}$ gives rise to a dual instrument
${\mathfrak I}^\ast:{\mathcal N}\times B({\mathcal H})^\ast\longrightarrow B({\mathcal H})^\ast$
defined by
\begin{equation}\label{M2}
 {\mathfrak I}^\ast(n)(X):=  {\mathfrak I}(n)^\ast(X)
\end{equation}
for all $n\in {\mathcal N}$ and $X\in B({\mathcal H})^\ast$. The condition that the total operation (\ref{OI4}) will
be trace-preserving translates into
\begin{equation}\label{M3}
 {\mathfrak I}^\ast({\mathcal N})({\mathbbm 1})=\sum_{n\in{\mathcal N}} {\mathfrak I}^\ast(n)({\mathbbm 1})
 =\sum_{n\in{\mathcal N}} \sum_{i\in{\mathcal I}_n} A_{ni}^\ast\, A_{ni}
 ={\mathbbm 1}
 \;.
\end{equation}
Thus every dual instrument yields a resolution of the identity by means of effects
\begin{equation}\label{eff1}
F_n:= {\mathfrak I}^\ast(n)({\mathbbm 1})=\sum_{i\in{\mathcal I}_n} A_{ni}^\ast\, A_{ni}
\;,
\end{equation}
and hence to a generalized observable in the sense of a {\em positive operator-valued measure}
$F=\left( F_n\right)_{n\in{\mathcal N}}$, see \cite{BLPY16}.Note, however, that compared to the general
definition in \cite{BLPY16} we will have to consider generalized observables only in the discrete, finite-dimensional case.
The traditional notion of ``sharp"  observables represented by self-adjoint operators corresponds to the special case
of a projection-valued measure $\left(P_n\right)_{n\in {\mathcal N}}$ satisfying $\sum_{n\in{\mathcal N}}P_n={\mathbbm 1}$.
From now on, by "observables" we always want to understand the generalized case.

It can be shown \cite{S20} that ``Maxwell instruments" are just pure instruments corresponding to sharp observables.
The example of imperfect erasure of qubits considered in Section \ref{sec:ER} suggest that the class of ``Maxwell instruments"
is too narrow to describe realistic conditional actions. First, imperfect erasure cannot be described by a pure instrument,
since the initial state of the ``heat bath" is not pure. Moreover, the  measurement of a sharp ``heat bath" observable does not
give rise to a sharp qubit observable. Hence it seems sensible to use general instruments to describe conditional action.
Fortunately, the main results on the entropy balance of conditional action in
\cite{S20} can be easily generalized to general instruments.

To this end we reconsider the map $\mathfrak{I}$ defined in (\ref{inst1}) by means of coupling the system $\Sigma$ to some environment $E$.
Recall that the environment $E$ is described by some Hilbert space ${\mathcal K}$ and an initial state $\sigma\in B_1^+\left({\mathcal K} \right)$.
Moreover, $V$ denotes the unitary time evolution of the total system and $\left(Q_n\right)_{n\in{\mathcal N}}$ a sharp environment observable.
It can be shown that (i) (\ref{inst1}) defines an instrument in the above sense and (ii) every instrument can be obtained
in this way, see Theorem 7.~14 of  \cite{BLPY16},  Exercise 8.~9 of \cite{NC00}, or Appendix \ref{sec:E}.
The special instrument defined in (\ref{inst1}) will be referred to as a ``measurement dilation" $\mathfrak{D}_{{\mathcal K,\sigma,V,Q}}$
of a given instrument $\mathfrak{I}$. If the initial state $\sigma$ of the environment is pure, $\sigma= P_\phi$, the measurement dilation
will also be denoted by  $\mathfrak{D}_{{\mathcal K,\phi,V,Q}}$. A measurement dilation of a given instrument $\mathfrak{I}$ is hence a physical
realization of $\mathfrak{I}$ by a L\"uders instrument of the extended system $\Sigma +E$ and a subsequent reduction to $\Sigma$.

Let a conditional action be described by the instrument $\mathfrak{I}$ with measurement dilation $\mathfrak{I}=\mathfrak{D}_{{\mathcal K,\sigma,V,Q}}$.
W.~r.~t.~this measurement dilation we define the two reduced states
\begin{equation}\label{rho1def}
\rho_1:= \mbox{Tr}_{\mathcal K}\left( \sum_{n\in{\mathcal N}}\left( {\mathbbm 1}\otimes Q_n\right)V\left(\rho\otimes \sigma \right)V^\ast
\left( {\mathbbm 1}\otimes Q_n\right) \right)
\;,
\end{equation}
and
\begin{equation}\label{rho2def}
\rho_2:= \mbox{Tr}_{\mathcal H}\left( \sum_{n\in{\mathcal N}}\left( {\mathbbm 1}\otimes Q_n\right)V\left(\rho\otimes \sigma \right)V^\ast
\left( {\mathbbm 1}\otimes Q_n\right) \right)
\;.
\end{equation}
Then the analogous arguments leading to the entropy balance (\ref{open3}) also prove:
\begin{prop}\label{prop0}
Under the preceding conditions the following holds:
\begin{equation}\label{entropy balance}
 \Delta S:= S(\rho)-S(\rho_1) \le S(\rho_2) -S(\sigma)
 \;.
\end{equation}
\end{prop}

In connection with the Szilard principle discussed in the next Section the following proposition will be of some interest:
\begin{prop}\label{prop1}
  The total operation $\mathfrak{I}({\mathcal N})$ of an instrument with measurement dilation
  $\mathfrak{I}=\mathfrak{D}_{{\mathcal K,\sigma,V,Q}}$ is independent of the environment observable $Q$.
\end{prop}
This means in particular that $\mathfrak{I}({\mathcal N})$ could even be realized by a coupling of $\Sigma$ to some
environment $E$, unitary time evolution and final state reduction, without any measurement at all.
The proof of Proposition \ref{prop1} can be found in Appendix \ref{sec:PR1}.

\section{The Szilard principle revisited}\label{sec:SP}

As mentioned in the Introduction, the ideas of L. Szilard \cite{S29} to resolve the apparent contradiction
between the results of the ``intervention of intelligent beings" and the second law are published more than nine decades ago
and are confined to classical physics. Therefore, the reconstruction of ``Szilard's principle" for quantum mechanics could appear as somewhat daring.
In this section, nevertheless, we will reconsider what we understand by ``Szilard's principle'' from the point of view developed in the present article.

According to this principle the entropy decrease of the system is compensated by the entropy costs of acquiring information about the
system's state. Recall that we have considered a so-called measurement dilation of the instrument $\mathfrak{I}$ describing state changes due to
conditional action which extends the system $\Sigma$ by an auxiliary system $E$ (environment).
For the ``standard dilation" given in Appendix \ref{sec:E} the dimension of the Hilbert space ${\mathcal K}$
corresponding to the auxiliary system $E$ equals the number of outcomes
$\left| {\mathcal N}\right|$ of the L\"uders measurement of $Q$ if the instrument $\mathfrak{I}$
is pure. It is therefore tempting to consider the auxiliary system $E$ as a ``memory'' that holds the information about the result
of the measurement and to interpret the ``entropy cost of information acquisition''
as the entropy $S(\rho_2)$ of the final state $\rho_2$ of $E$ after the measurement.
The probabilities $p_n,\,n\in {\mathcal N},$ of the various outcomes are given by $p_n=\mbox{Tr} \left( \rho\,F_n\right)$,
where $F=\left( F_n\right)_{n\in {\mathcal N}}$ is the observable (\ref{eff1}) corresponding to the conditional action
and $\rho$ is the initial state of the system $\Sigma$. The Shannon entropy $H(p)$ of the probability distribution
$\left(p_n\right)_{n\in {\mathcal N}}$
is independent of any measurement dilation and will be called the ``Shannon entropy of the experiment"
\begin{equation}\label{ShannonExp}
  H(\rho,F):=H(p):=-\sum_{n\in {\mathcal N}} p_n\, \log p_n
  \;.
\end{equation}

Then we can prove the following inequality, which confirms Szilard's principle in the above given version:

\begin{theorem}\label{T1}{\em (Szilard's principle - quantum case)}\\
The entropy decrease $\Delta S = S(\rho)-S(\rho_1)$ of a conditional action corresponding to a pure instrument
$\mathfrak{I}$ is bounded by the Shannon entropy of the experiment, i.~e.,
\begin{equation}\label{theorem1}
  \Delta S \le H(\rho,F)
  \;.
\end{equation}
\end{theorem}
For the proof see Appendix \ref{sec:PR2}.
It is worth noting that the bound in (\ref{theorem1}) is independent of the pure instrument
describing the conditional action and depends only on the probabilities $p_n= \mbox{Tr} \left(\rho F_n\right)$.
The theorem is trivially satisfied if the conditional action leads to an {\em increase} of entropy, i.~e., $\Delta S\le 0$,
as in the case of a L\"uders measurement without any conditional action.
Another trivial case is given if the observable $F$ is sharp and the projections $F_n$ are one-dimensional.
In this case $S(\rho)\le S\left( \sum_n F_n \rho F_n\right)=H(\rho,F)$
and the theorem holds since $S(\rho_1)\ge 0$. In other words: Entropy cannot fall below the value of zero.

Otherwise the bound (\ref{theorem1}) is non-trivial. Consider the example of ${\mathcal H}={\mathbb C}^3$ with three
mutually orthogonal one-dimensional projections $P_1,P_2,P_3$,
\begin{equation}\label{ex1}
  \rho={\textstyle \frac{1}{2}}P_1+{\textstyle \frac{3}{10}}P_2+{\textstyle \frac{1}{5}}P_3
  \;,
\end{equation}
and $F=(P_1,P_2+P_3)$. It follows that
\begin{equation}\label{ex2}
  S(\rho)=-{\textstyle \frac{1}{2}} \log \left({\textstyle \frac{1}{2}}\right)-{\textstyle \frac{3}{10}}
   \log \left({\textstyle \frac{3}{10}}\right)-{\textstyle \frac{1}{5}}
   \log \left({\textstyle \frac{1}{5}}\right)\approx 1.02965 > H(\rho,F)=\log 2 \approx 0.693147
   \;,
\end{equation}
and hence, in this example, Theorem \ref{T1} says more than just that the entropy of $\rho$ cannot drop to negative values.
It is straightforward to construct a Maxwell instrument corresponding to the observable $F$ such that
$\rho_1= {\textstyle \frac{4}{5}}\,P_1+{\textstyle \frac{1}{5}}\,P_2$ and hence $\Delta S\approx 0.529251<\log 2 \approx 0.693147$
in accordance with Theorem \ref{T1}.\\

A slight generalization of Theorem \ref{T1} is the following:

\begin{cor}\label{Cor1}
 The upper bound (\ref{theorem1}) also holds if the instrument $\mathfrak{I}$ can be written as a convex linear combination of pure instruments
 with the same set of outcomes ${\mathcal N}$.
\end{cor}
For the proof see Appendix \ref{sec:PR3}.
A pure instrument has a standard dilation with one-dimensional projections $Q_n, \; n\in {\mathcal N},$ and
a pure initial state $\sigma=P_\phi$ of $E$. If we extend this standard dilation by considering a real mixed initial state
$\sigma$ of $E$ we obtain a convex combination of pure instruments for which  Corollary  \ref{Cor1} holds.
But not every instrument is a convex linear combination of pure ones. Actually, there exist instruments where
the bound of entropy decrease given in (\ref{theorem1}) is violated.

To provide an example we consider the (perfect) erasure of two qubits.
Thus ${\mathcal H}={\mathbbm C}^2\otimes {\mathbbm C}^2 \cong {\mathbbm C}^4$ and we consider an orthonormal basis of
${\mathcal H}$ denoted by
$\left(\psi_1=\uparrow\uparrow,\,\psi_2=\uparrow\downarrow,\,\psi_3=\downarrow\uparrow,\,\psi_4=\downarrow\downarrow\right)$.
The conditional action maps all these four basis states onto the default state $\psi_4=\downarrow\downarrow$.
It has the following measurement dilation: ${\mathcal  K}={\mathcal H}$, initial auxiliary state $\phi=\downarrow\downarrow$,
unitary time evolution $V$ of the total system defined by
$V(\Phi\otimes \Psi)=\Psi\otimes \Phi$ and L\"uders measurement of the auxiliary observable
$Q=\left(Q_\nu\right)_{\nu=1,\ldots,4}=\left(|\psi_\nu\rangle \langle \psi_\nu|\right)_{\nu=1,\ldots,4}$.
The corresponding instrument $\mathfrak{I}=\mathfrak{D}_{{\mathcal  K},\phi,V,Q}$ is pure and hence  satisfies (\ref{theorem1}).

Then we consider another instrument $\widetilde{\mathfrak{I}}$ by changing the auxiliary observable to
$\widetilde{Q}=\left(Q_1+Q_2 ,Q_3+Q_4 \right)=\left(|\psi_1\rangle\langle\psi_1 | \otimes {\mathbbm 1} ,|\psi_2\rangle \langle\psi_2 | \otimes {\mathbbm 1} \right)$.
The corresponding system observable
$\widetilde{F}=\left( |\uparrow\rangle \langle \uparrow|\otimes {\mathbbm 1},|\downarrow\rangle \langle \downarrow|\otimes {\mathbbm 1}\right)$
is also two-valued and corresponds to a measurement of the first qubit w.~r.~t.~the considered basis.
All other components of $\mathfrak{D}_{{\mathcal  K},\phi,V,Q}$ are left unchanged.
Consider the initial state $\rho={\textstyle \frac{1}{4}}{\mathbbm 1}$ of the system with $S(\rho)=\log 4$
and $H(\rho,\widetilde{F})=\log 2$. It follows that $V(\rho \otimes P_\phi)V^\ast=P_\phi \otimes \rho$ and hence
$\rho_1= P_\phi$ and $S(\rho_1)=0$. Consequently, $\Delta S= S(\rho)-S(\rho_1)=\log 4 >\log 2=H(\rho,\widetilde{F})$ 
in contrast to (\ref{theorem1}).

Similar examples abound: Whenever the entropy decrease $\Delta S$ due to a conditional action is larger than $\log 2$ then a corresponding measurement
dilation can be modified to yield $S(\rho,\widetilde{F})=\log 2$ without changing $\Delta S$ due to Proposition \ref{prop1}.
The modified instrument  $\widetilde{\mathfrak{I}}$ cannot be written as a convex linear combination of pure instruments according to
Corollary \ref{Cor1}.

As a conclusion for the evaluation of Szilard's principle we can state that there are
examples of conditional actions where the entropy decrease in the system can be explained
by an entropy increase at least as large in a memory, as well as counter examples.
In the counterexamples, however, we have no violation of the second law,
but only an impossibility to reduce the auxiliary system to its function as a memory.
This is especially true for the limiting case of an entropy reduction without measurement.

The example of Section \ref{sec:ER}, see Figure \ref{FIGSS8},  
shows that the upper bound (\ref{theorem1}) of the entropy decrease
holds for a larger class of conditional actions than given by Theorem \ref{T1} or Corollary \ref{Cor1}.
It remains an open task to determine this class more precisely.

\section{Connections to the OLR approach}\label{sec:OLR}

There exists a vast amount of literature on Maxwell's demon and related questions \cite{footnote}.
Among them is an article that comes rather close to the results of the present work, namely \cite{A13},
that deals with Szilard's engine and where we read in the abstract:

\begin{quote}
In this paper, Maxwell's Demon is analyzed within
a ``referential" approach to physical information that defines and quantifies the Demon's
information via correlations between the joint physical state of the confined molecule and
that of the Demon's memory. On this view [...] information is erased not during the memory reset
step of the Demon's cycle, but rather during the expansion step, when these correlations are
destroyed.
\end{quote}

The mentioned notion of ``observer-local referential (OLR) information" is further outlined in \cite{A17}.
A detailed comparison of the ``conditional action approach" and the ``OLR approach" is beyond the scope of this paper.
Arguably, a key difference is that we could not model the formation of a correlation
by a measurement and the subsequent destruction of that correlation by a conditional action
as a sequence of state changes and instead had to use measurement dilation as a surrogate, see Section \ref{sec:SU}.

To illustrate the nevertheless existing connections between the two approaches,
we will derive another upper bound for the entropy decrease analogous to Theorem \ref{T1}
using the notion of OLR information. We will restrict ourselves to the case of
conditional actions described by Maxwell instruments $\mathfrak{M}$, i.~e., instruments  of the form (\ref{condact1}).
Let $\mathfrak{L}$ denote the corresponding L\"uders instrument  of the form (\ref{mess1}) such that
$\mathfrak{M}$ and $\mathfrak{L}$ share the same Hilbert space ${\mathcal H}$ and the same sharp observable
$P=\left( P_n\right)_{n\in{\mathcal N}}$. Further consider the standard measurement dilations
\begin{equation}\label{dil1}
 \mathfrak{L}=\mathfrak{D}_{{\mathcal K},\phi',V',Q'},\quad\mbox{and} \quad \mathfrak{M}=\mathfrak{D}_{{\mathcal K},\phi,V,Q}
 \;,
\end{equation}
as explicitly constructed in Appendix \ref{sec:E} but specialized to the case of pure instruments, see also \cite{S20}.
W.~r.~t.~ these measurement dilations we further define
\begin{eqnarray}
\label{rho12a}
 \rho_{12} &:=& \sum_{n\in{\mathcal N}} \left( \mathbbm{1}-Q_n\right)\,V\,\left(\rho \otimes P_\phi\right)\,V^\ast \,\left( \mathbbm{1}-Q_n\right)\,,\\
 \label{rho2b}
  \rho_1&:=&\mbox{Tr}_{\mathcal K} \, \rho_{12},\quad \mbox{and}\quad\rho_2 := \mbox{Tr}_{\mathcal H} \, \rho_{12}
  \;,
\end{eqnarray}
and analogously for the primed quantities:
\begin{eqnarray}
\label{rho12c}
 \rho_{12}' &:=& \sum_{n\in{\mathcal N}} \left( \mathbbm{1}-Q'_n\right)\,V'\,\left(\rho \otimes P_{\phi'}\right)\,V'^\ast \,\left( \mathbbm{1}-Q'_n\right)\,,\\
 \label{rho12d}
  \rho_1'&:=&\mbox{Tr}_{\mathcal K} \, \rho_{12}',\quad \mbox{and}\quad\rho_2' := \mbox{Tr}_{\mathcal H} \, \rho_{12}'
  \;.
\end{eqnarray}
In accordance with \cite{A17} we define the OLR information
\begin{eqnarray}
 \label{OLRa}
  {\mathcal I} &:=& S(\rho_{1})+S(\rho_{2}) -S(\rho_{12})=: S_1+S_2-S_{12}\;,\\
  \label{OLRb}
  {\mathcal I}' &:=& S(\rho_{1}')+S(\rho_{2}') -S(\rho_{12}')=: S_1'+S_2'-S_{12}'\;,\\
  \label{OLRc}
  \Delta {\mathcal I}&:=&  {\mathcal I}'- {\mathcal I}
  \;.
\end{eqnarray}
Further, let
\begin{equation}\label{S0def}
  S_0:= S(\rho)=S(\rho \otimes P_\phi)=S(\rho \otimes P_{\phi'})
  \;,
\end{equation}
and
\begin{equation}\label{DeltaSdef}
  \Delta S:= S_0-S_1
  \;.
\end{equation}
Then we can prove the following
\begin{prop}\label{prop2}
  Under the preceding conditions the entropy decrease $\Delta S$ is bounded from above by
  \begin{equation}\label{DSI}
   \Delta S \le \Delta {\mathcal I}
   \;.
  \end{equation}
\end{prop}
The proof of Proposition \ref{prop2} can be found in Appendix \ref{sec:PR4}.
Of course, this is only a first step to analyze the mentioned relations,
since the conditions of Proposition \ref{prop2} are rather limited,
e.~g., by the fact that only the standard dilation is considered and not an arbitrary measurement dilation.

\section{Imperfect erasure of a qubit}
\label{sec:ER}
The role of the detailed example considered in this Section is twofold:
First, we can explain and illustrate the definitions of the previous sections using a non-trivial but still computable example.
Second, this reasonably realistic case also demonstrates the viability of the general theory.

\subsection{Definition of the model}
\label{sec:ERD}

\begin{figure}[t]
\centering
\includegraphics[width=0.7\linewidth]{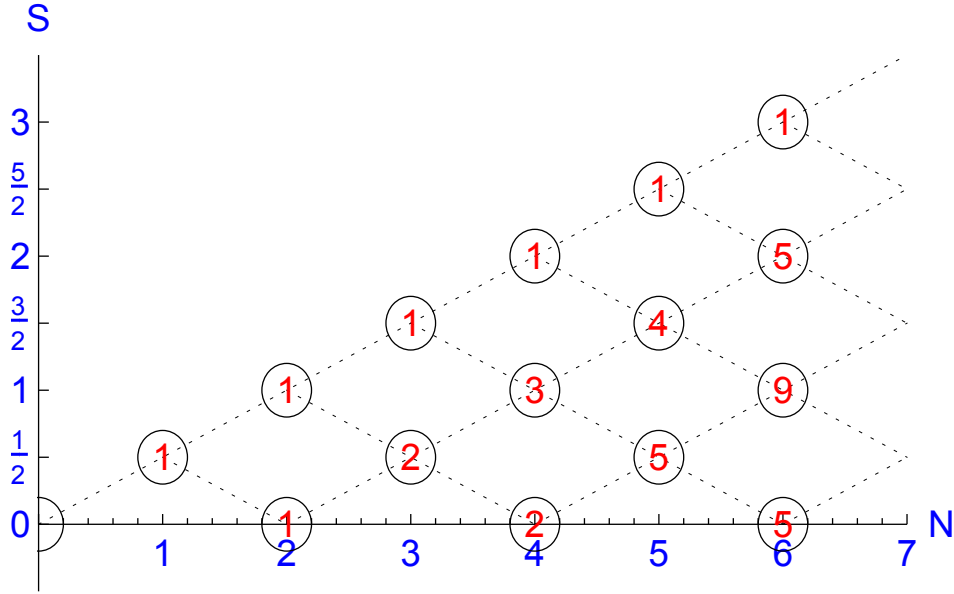}
\caption{``Half Galton Board": The degeneracies  $D_N(S)$ of states $\left| S,M\right\rangle$ generated by coupling $N$ spins with $s=1/2$
can be obtained by the (red) number of coupling paths that start at $(N=0,S=0)$ and terminate at $(N,S)$.
}
\label{FIGRW1}
\end{figure}

We consider a system of $N$ spins with spin quantum number $s=1/2$ equipped with a uniform anti-ferromagnetic Heisenberg coupling
and a Zeeman term. This leads to a Hamiltonian
\begin{equation}\label{Ham1}
  H_N=J\,\sum_{1\le\mu<\nu\le N}\op{\mathbf{s}}_\mu\cdot\op{\mathbf{s}}_\nu \,+ B\sum_{\mu=1}^N \op{s}_\mu^z
  \;,
\end{equation}
where $J>0$ and $B>0$ are dimensionless physical parameter characterizing the spin system.
$\op{\mathbf{s}}_\mu=\left( \op{s}_\mu^x,\op{s}_\mu^y,\op{s}_\mu^z\right)$
represents the vector of spin operators at the site $\mu$.
It is well-known that the corresponding time evolution can be analytically calculated
since we may write the Hamiltonian in the form
\begin{equation}\label{Ham2}
 H_N=\frac{J}{2}\left(\op{S}^2-\frac{3 N}{4} \mathbbm{1}\right) + B\,\op{S}^z
 \;,
\end{equation}
where $\op{\mathbf{S}}:=\sum_{\mu=1}^N \op{\mathbf{s}}_\mu$ denotes the total vector of spin operators and $\op{S}^z$ its $z$-component.
Since $\op{S}^2$ and $\op{S}^z$ commute they possess a system of common eigenvectors
$\left| \alpha; S,M\right\rangle$ satisfying the eigenvalue equations
\begin{eqnarray}
\label{eig1}
  \op{S}^2\,\left| \alpha;S,M\right\rangle &=& S(S+1)\,\left| \alpha;S,M\right\rangle\;, \\
  \label{eig2}
   \op{S}^z\,\left| \alpha;S,M\right\rangle &=& M\,\left| \alpha;S,M\right\rangle
   \;,
\end{eqnarray}
and hence
\begin{equation}\label{eigHam}
  H_N\,\left| \alpha;S,M\right\rangle =
  \left({\textstyle\frac{J}{2}}\left(S(S+1)-{\textstyle \frac{3N}{4}}\right)+B\,M \right)\left| \alpha;S,M\right\rangle
 =: E_N(S,M)\,\left|\alpha;S,M\right\rangle
  \;.
\end{equation}
The theory of coupling angular momenta treated in many textbooks yields that the quantum number $S$
assumes the values $\frac{1}{2},1,\frac{3}{2},\ldots,\frac{N}{2}$ for odd $N$ and
$0,1,2,\ldots,\frac{N}{2}$ for even $N$ and $M=-S,-S+1,\ldots, S-1,S$. The symbol ``$\alpha$" stands for
further quantum numbers that allow for the degeneracy $D_N(S)$ of the eigenspaces with common eigenvalues $S(S+1)$ and $M$
of $\op{S}^2$ and $\op{S}^z$, resp., such that the normalized eigenvectors $\left| \alpha;S,M\right\rangle$
will be unique up to a phase. For a selection of such degeneracies see Figure \ref{FIGRW1}.

We will consider a single spin with $s=1/2$  with Hilbert space ${\mathcal H}\cong {\mathbbm C}^2$
representing one qubit and try to realize the erasure of the qubit
by  coupling the single spin to a ``heat bath" consisting of $N=6$ uniformly coupled spins
such that the total Hamiltonian $H$ will be
of the form $H_7$.
Moreover, we  choose $J=B=1$ thereby fixing a natural energy unit and corresponding physical units
of time and temperature by setting $\hbar=k_B=1$. The choice of the ``heat bath" with $N=6$ spins 
has the pleasant consequence that all relevant quantities can be directly calculated 
by the means of computer-algebraic means without resorting to the theory
of coupling angular momenta.

The ``heat bath" with Hilbert space ${\mathcal K}\cong {\mathbbm C}^{64}$
has a ground state with energy $E_6(0,0)=E_6(1,-1)=-\frac{9}{4}$ that is $14$-fold degenerate.
This follows from the degeneracies $D_6(S=0)=5$ and $D_6(S=1)=9$, see Figure \ref{FIGRW1}.
Let $Q_0$ denote the projector onto the  corresponding eigenspace and $Q_1$ the complementary projector
such that $Q_0+Q_1={\mathbbm 1}_{\mathcal K}$.

We will assume that initially the ``heat bath" is in its ground state $\sigma:={\textstyle \frac{1}{14}} Q_0$
corresponding to the temperature $T=0$
whereas the single spin is in an arbitrary mixed state $\rho$. Then a unitary time evolution
$U_t:= \exp\left( -{\sf i} \,t \,H \right)$ takes place followed by a L\"uders measurement of the sharp heat bath observable $(Q_0,Q_1)$.
After this measurement we consider the two reduced states
\begin{equation}\label{rho1}
\rho_1= \mbox{Tr}_{\mathcal K}\left( \sum_{n=0,1}\left( {\mathbbm 1}\otimes Q_n\right)U_t\left(\rho\otimes \sigma \right)U_t^\ast
\left( {\mathbbm 1}\otimes Q_n\right) \right)
\;,
\end{equation}
and
\begin{equation}\label{rho2}
\rho_2= \mbox{Tr}_{\mathcal H}\left( \sum_{n=0,1}\left( {\mathbbm 1}\otimes Q_n\right)U_t\left(\rho\otimes \sigma \right)U_t^\ast
\left( {\mathbbm 1}\otimes Q_n\right) \right)
\;.
\end{equation}
Obviously, $\rho_1$ is the result of the total operation $\rho_1=\mathfrak{I}({\mathcal N})(\rho)$ corresponding
to the instrument
\begin{equation}\label{instr}
\mathfrak{I}(n)(\rho):=\mbox{Tr}_{\mathcal K}\left(\left( {\mathbbm 1}\otimes Q_n\right)U_t\left(\rho\otimes \sigma \right)U_t^\ast
\left( {\mathbbm 1}\otimes Q_n\right) \right)
\;,
\end{equation}
where $n\in{\mathcal N}= \{0,1\}$.

It turns out that for the special model we have considered the matrix elements of $\rho_n$
are $4\pi$-periodic functions of $t$. Instead of dwelling into a debate how to cope with these oscillating terms
we simply make the choice $t=2\pi$, i.e., we consider the time evolution of a half period before performing the final measurement.
This choice gives reasonable results which suffices to constructing an example of the general theory outlined in this paper.
Now all parameters of our model for imperfect erasure are fixed and we proceed by presenting the relevant results
without explicating the further details of the computer-algebraic calculation.

\subsection{Results on the instrument $\mathfrak{I}$}
\label{sec:ERI}

The first results concern the calculation and visualization of the total trace-preserving
operation $\rho \mapsto \rho_1=\mathfrak{I}({\mathcal N})(\rho)$.
Recall that $\rho\in {\mathcal B}({\mathcal H})$ where the latter is a $4$-dimensional space spanned by the four
Pauli matrices
\begin{equation}\label{basisB}
  \sigma_0= \left( \begin{array}{cc}
                     1& 0 \\
                     0 & 1
                   \end{array}\right),\;
  \sigma_1= \left( \begin{array}{cc}
                     0& 1 \\
                     1 &0
                   \end{array}\right),\;
  \sigma_2= \left( \begin{array}{cc}
                     0& -{\sf i} \\
                     {\sf i} & 0
                   \end{array}\right),\;
  \sigma_3= \left( \begin{array}{cc}
                     1& 0 \\
                     0 & -1
                   \end{array}\right)
  \;.
\end{equation}
They are mutually orthogonal w.~r.~to the Euclidean scalar product $(A,B)\mapsto \mbox{Tr}\left( A B\right)$ of Hermitean $2\times 2$-matrices
and have the length $\sqrt{2}$. W.~r.~t.~this basis the  total operation $\mathfrak{I}({\mathcal N})$ can be represented
by the $4\times 4$ matrix
\begin{equation}\label{totalMatrix}
 {\mathbf I}={\textstyle \frac{1}{7}}
 \left(
\begin{array}{cccc}
 7 & 0 & 0 & 0 \\
 0 & 1 & 0 & 0 \\
 0 & 0 & 1 & 0 \\
 -4 & 0 & 0 & 3 \\
\end{array}
\right)
\;.
\end{equation}
Note that $\mathfrak{I}({\mathcal N})$ being trace preserving is equivalent to the property $\mathfrak{I}^\ast({\mathcal N})(\sigma_0)=\sigma_0$
of the dual instrument $\mathfrak{I}^\ast$. The latter is represented by the transposed matrix ${\mathbf I}^\top$ and hence
the first row of ${\mathbf I}$,  that corresponds to the first column of ${\mathbf I}^\top$, must be necessarily of the form $(1,0,0,0)$.

The density matrices $\rho$ in the Hilbert space
${\mathcal H}\cong {\mathbbm C}^2$ can be represented by the points $(x_1,x_2,x_3)^\top$
of a unit ball in ${\mathbbm R}^3$ such that the pure states corresponding to
one-dimensional projectors form its surface ${\mathcal S}^2$, the so-called  ``Bloch sphere".
This representation is given by
\begin{equation}\label{Bloch1}
  \rho= {\textstyle\frac{1}{2}} \left( \sigma_0 + \sum_{i=1}^{3}x_i\,\sigma_i\right)
  \;.
\end{equation}

\begin{figure}[t]
\centering
\includegraphics[width=0.7\linewidth]{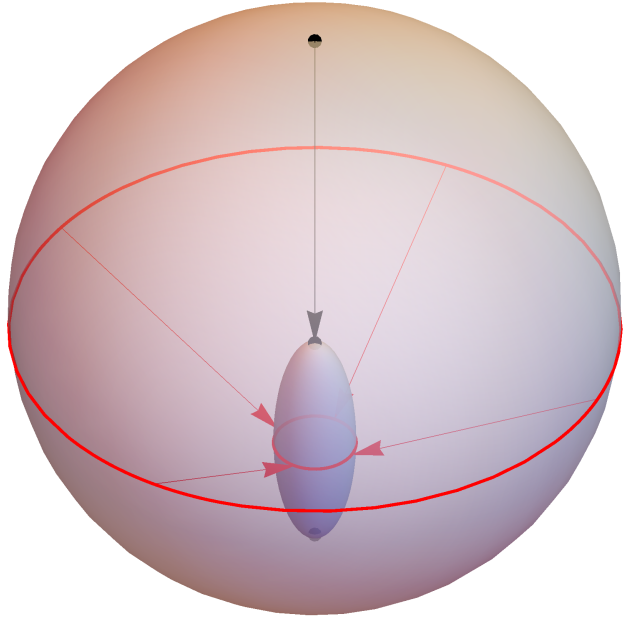}
\caption{Visualization of the total operation $\mathfrak{I}({\mathcal N})$
of imperfect erasure of a qubit as the affine mapping of the Bloch sphere ${\mathcal S}^2$
onto an ellipsoid ${\mathcal E}$ that touches ${\mathcal S}^2$ at the south pole.
}
\label{FIGJ7}
\end{figure}

Under the total operation $\mathfrak{I}({\mathcal N})$,
which is an affine map on states, the Bloch sphere is mapped onto an ellipsoid ${\mathcal E}$ lying inside ${\mathcal S}^2$, see Figure \ref{FIGJ7}.
Due to
$\det({\mathbf I})={\textstyle\frac{1}{7}}\times {\textstyle\frac{1}{7}}\times {\textstyle\frac{3}{7}}={\textstyle\frac{3}{343}}$
the volume of ${\mathcal S}^2$ is compressed to less than one percent.
This volume compression is typical for conditional action.
The invariance of ${\mathcal E}$ under rotations about the $3$-axis is due to the azimuthal symmetry of the Hamiltonian (\ref{Ham1})
and of the initial state of the ``heat bath".

Some properties of the mapping  $\mathfrak{I}({\mathcal N})$ can be read off the matrix (\ref{totalMatrix}):
The south pole of ${\mathcal S}^2$ is mapped onto itself
and this is the only point where ${\mathcal E}$ touches the Bloch sphere. Since this south pole corresponds to the state
$\rho= {\textstyle \frac{1}{2}}\left( \sigma_0- \sigma_3\right)$ its invariance under $\mathfrak{I}({\mathcal N})$ is reflected by
the equation $\mathbf{I}(1,0,0,-1)^\top=(1,0,0,-1)^\top$. Physically, the south pole represents the ground state of the qubit corresponding
to the Gibbs state with temperature $T=0$. Its invariance under  $\mathfrak{I}({\mathcal N})$ hence means that the quit remains in its
ground state if it is coupled to a ``heat bath" of temperature $T=0$, which is very plausible.

The orientation relative to the coordinate system and the semi-axes $({\textstyle\frac{1}{7}},
{\textstyle\frac{1}{7}},{\textstyle\frac{3}{7}})$ of the ellipsoid ${\mathcal E}$
can be read off the lower right $3\times 3$-submatrix of ${\mathbf I}$, see (\ref{totalMatrix}).
The center of ${\mathcal E}$ lies at $x_3=-{\textstyle\frac{4}{7}}$ corresponding to the state
\begin{equation}\label{center}
 \rho_1'= {\textstyle \frac{1}{2}}\left( \sigma_0- {\textstyle\frac{4}{7}}\sigma_3\right)=
  {\textstyle \frac{1}{14}}\left(
\begin{array}{cc}
 3 & 0 \\
 0 & 11 \\
\end{array}
\right)
\;,
\end{equation}
and the north pole of ${\mathcal S}^2$ is mapped onto
\begin{equation}\label{north}
 \rho_1''= {\textstyle \frac{1}{2}}\left( \sigma_0- {\textstyle\frac{1}{7}}\sigma_3\right)=
 {\textstyle \frac{1}{7}}\left(
\begin{array}{cc}
 3 & 0 \\
 0 & 4 \\
\end{array}
\right)
\;.
\end{equation}
By a perfect erasure of a qubit the Bloch sphere would be completely mapped onto the south pole; a more realistic scenario
corresponds to a mapping onto a small ellipsoid close to the south pole. The present example may not yield the best possible result;
however its virtue lies in the fact that ${\mathcal E}$  can be analytically calculated and is rather simple in form.\\

After having analyzed the total operation $\mathfrak{I}({\mathcal N})$ we proceed by considering the two components
$\mathfrak{I}(n),\; n=0,1,$ of the instrument $\mathfrak{I}$. We will determine the corresponding Kraus operators $A_{nmj}$
such that
\begin{equation}\label{Kraus2}
 \mathfrak{I}(n)(\rho)= \sum_{mj}A_{nmj}\,\rho\, A_{nmj}^\ast,\quad n\in{\mathcal N}
 \;.
\end{equation}
The index  $i$ occurring in (\ref{OI3}) has  been replaced here by a multi-index $i=(m,j)$.
According to the general theory the Kraus operators $A_{nmj}$
can be derived from the measurement dilation $\mathfrak{I}={\mathfrak D}_{{\mathcal K},\sigma,V,Q}$
by means of the equation
\begin{equation}\label{Kraus1}
  \left\langle a \right| A_{nmj} \left| b \right\rangle =
   \sqrt{q_j}  \left\langle a \otimes \phi_m \right| V \left| b \otimes \psi_j \right\rangle,\quad a,b\in {\mathcal H},\;
   m\in{\mathcal M}_n
   \;,
\end{equation}
see \cite{NC00}, 8.35.
Here we have used the spectral decomposition of the initial state $\sigma$ of the auxiliary system
\begin{equation}\label{tauspectral}
 \sigma=\sum_j q_j\, \left| \psi_j \right\rangle \left\langle \psi_j \right|
 \;,
\end{equation}
and that of the projector $Q_n$
\begin{equation}\label{Qnspectral }
 Q_n= \sum_{m\in{\mathcal M}_n} \left| \phi_m \right\rangle \left\langle \phi_m \right|
\;.
\end{equation}
The latter is defined w.~r.~t.~a suitable partition ${\mathcal M}=\biguplus_n {\mathcal M}_n$ of the index set ${\mathcal M}$
corresponding to an orthonormal basis $\left( \phi_m\right)_{m\in{\mathcal M}}$ of ${\mathcal K}$ adapted to
the sharp observable $\left( Q_n\right)_{n\in{\mathcal N}}$.

\begin{figure}[t]
\centering
\includegraphics[width=0.7\linewidth]{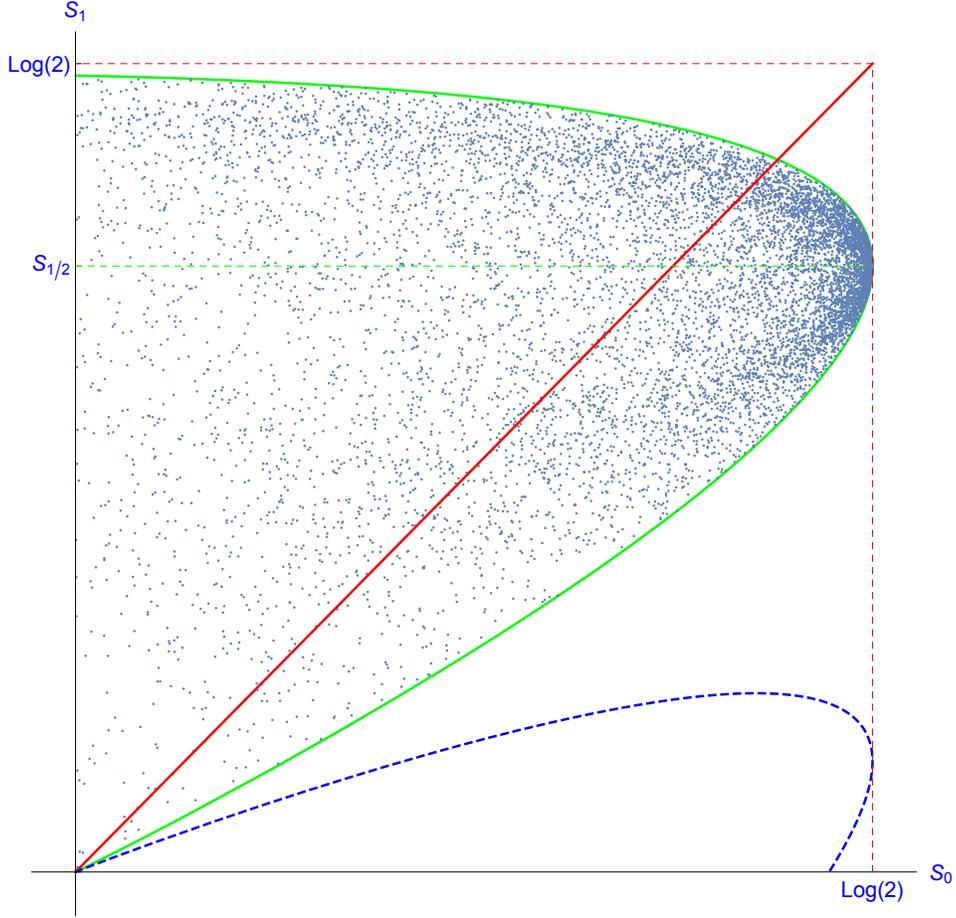}
\caption{Plot of the final entropy $S_1$ versus the initial one $S_0$ for an imperfect erasure of a qubit. The blue dots correspond to $10,000$ randomly
chosen initial states $\rho$ of the single spin. The enveloping green curve is analogously calculated for the one-dimensional
family
$\rho(p)$ according to (\ref{rhop1},\ref{rhop2}) and reaches its maximum of $S_0=\log 2$ for $p={\textstyle\frac{1}{2}}$ (dashed green line).
The corresponding value $S_{1/2}$ of $S_1$ is given by (\ref{Shalf}). Only for the points below the (red) line $S_0=S_1$ there occurs
a decrease of entropy due to the conditional action. For the green curve this will happen if $0<p<p_1$, where $p_1$ is given by (\ref{p1}).
The dashed blue curve represents $S_0-H(\rho(p),F)$ according to (\ref{ShannonExp}). Therefore, the bound  (\ref{theorem1}) holds although
the conditions of Theorem \ref{T1} or Corollary \ref{Cor1} are not satisfied.
}
\label{FIGSS8}
\end{figure}

For the present example it appears at first sight that we would need $14\times 64= 896$ Kraus operators.
Fortunately, only $94$ Kraus operators do not vanish.
Also, they can be combined and simplified so that only the following three operators remain:
\begin{equation}\label{Kraus3}
 A_1=\left(
\begin{array}{cc}
 \sqrt{\textstyle\frac{5}{14}} & 0 \\
 0 & \sqrt{\textstyle\frac{5}{14}} \\
\end{array}
\right),\quad
 A_2=\left(
\begin{array}{cc}
 -{\textstyle\frac{1}{\sqrt{14}}} & 0 \\
 0 & {\textstyle\frac{3}{\sqrt{14}}} \\
\end{array}
\right),\quad
A_3=\left(
\begin{array}{cc}
 0 & 0 \\
{\textstyle \frac{2}{\sqrt{7}}} & 0 \\
\end{array}
\right)
\;.
\end{equation}
Here the first two operators $A_1$ and $A_2$ belong to $\mathfrak{I}(0)$ and $A_3$ to $\mathfrak{I}(1)$.
Hence the observable $F=(F_0,F_1)$ given by the instrument $\mathfrak{I}$ is obtained as
\begin{equation}\label{obs12}
 F_0=A_1^\ast\,A_1+A_2^\ast\,A_2= \left(
\begin{array}{cc}
{ \textstyle\frac{3}{7}} & 0 \\
 0 & 1 \\
\end{array}
\right)
\quad \mbox{and} \quad
 F_1=A_3^\ast\,A_3= \left(
\begin{array}{cc}
 {\textstyle\frac{4}{7}} & 0 \\
 0 & 0 \\
\end{array}
\right)
\;,
\end{equation}
satisfying $F_0+F_1={\mathbbm 1}$, as it is required for $F$ being an observable.
The fact that $F$ is not a sharp observable means that, despite energy conservation,
there is no perfect correlation between the energy of the individual spin and that of the ``heat bath''.
The latter would be expected only on the basis of time-dependent perturbation theory
(Fermi's ``Golden Rule'') and does not hold for a finite interaction between the spin and the ``heat bath''.

\subsection{Results on the entropy balance}
\label{sec:ERE}

\begin{figure}[t]
\centering
\includegraphics[width=0.7\linewidth]{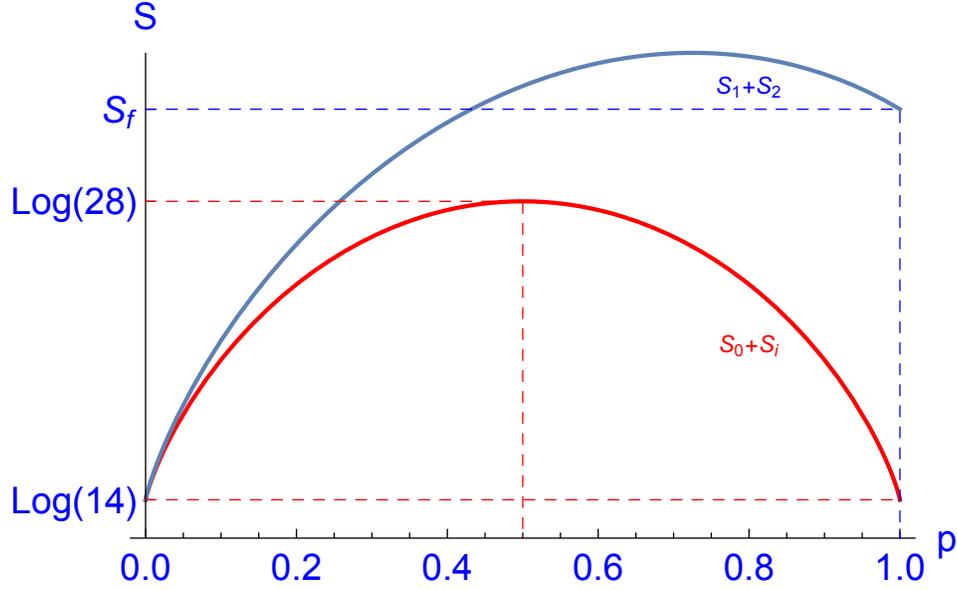}
\caption{Plot of the initial total entropy $S_0+S_i$  (red curve) and the final entropy $S_1+S_2$ (blue curve)
calculated for the one-parameter family $\rho(p)$ according to (\ref{rhop0}) such that $S_0=S(\rho(p))$.
Obviously, $S_0+S_i< S_1+S_2$ for $0<p\le 1$.
$S_i=\log 14$ denotes the initial
entropy of the ``heat bath" and hence $S_0+S_i$ assumes its maximum $\log 2 +\log 14 = \log 28\approx 3.3322$ at $p=1/2$.
At $p=1$ the final entropy $S_1+S_2$ assumes the value $S_f\approx 3.54621$ according to (\ref{Sf}).
}
\label{FIGSS6}
\end{figure}

We now turn to the entropy balance. First we plot $S_1:= S(\rho_1)$ versus $S_0:= S(\rho)$, see Figure \ref{FIGSS8}.
For the one-parameter family of states
\begin{equation}\label{rhop0}
 \rho(p):=
\left( \begin{array}{cc}
                   p & 0 \\
                    0 & 1-p
       \end{array}
      \right)
\;,
\end{equation}
where  $ 0\le p\le 1$, we obtain a curve with parametric representation
\begin{eqnarray}\label{rhop1}
S_0(\rho(p))&=& -p \log (p)-(1-p) \log (1-p),\\
\label{rhop2}
S_1(\rho(p))&=&{\textstyle\frac{1}{7}} \left((3 p-7) \log \left(1-{\textstyle\frac{3 p}{7}}\right)-3 p \log
   \left({\textstyle\frac{3 p}{7}}\right)\right)
   \;,
\end{eqnarray}
see the green curve in Figure \ref{FIGSS8}. The value $p=1/2$ corresponds to the maximum $\log 2$ of $S_0$ and
the value
\begin{equation}\label{Shalf}
S_{1/2}= {\textstyle\frac{1}{7}} \left({\textstyle\frac{11}{2}} \log \left({\textstyle\frac{14}{11}}\right)+{\textstyle\frac{3}{2}} \log
   \left({\textstyle\frac{14}{3}}\right)\right)
\end{equation}
of $S_1$ corresponding to the entropy of the center of the ellipsoid ${\mathcal E}$.
This curve is the envelope of the set of all points $\left(S_0(\rho) ,S_1(\rho) \right)$ as can be seen as follows.

The surfaces with constant entropy (``adiabatic surfaces") are the concentric spheres ${\mathcal S}$ inside the Bloch sphere
(together with the center considered as a degenerate sphere).
The set of states $\rho$ corresponding to such a concentric sphere ${\mathcal S}$
is mapped under $\mathfrak{I}({\mathcal N})$ onto an ellipsoid ${\mathcal E}'\subset {\mathcal E}$ that is also invariant
under rotations about the $3$-axis.
The north pole $N$ of ${\mathcal S}$ corresponding to the state $\rho(p)$ is mapped onto the north pole $N'$ of ${\mathcal E}'$.
Similarly, the south pole $S$ of  ${\mathcal S}$ corresponding to the state $\rho(1-p)$ is mapped onto the south pole $S'$ of ${\mathcal E}'$.
The total ellipsoid ${\mathcal E}'$ is bounded by the two concentric spheres through $N'$ and $S'$ and hence the entropy
of all states corresponding to ${\mathcal E}'$ is bounded by $S_1(\rho(p))$ and $S_1(\rho(1-p))$.

Decrease of entropy, i.~e., $S(\rho_1) < S(\rho_0)$, will not always occur. For example, the north pole of
${\mathcal S}^2$  corresponds to a pure state of vanishing entropy and is mapped onto a mixed state with positive entropy.
For the states $\rho(p)$ decrease of entropy is equivalent to
\begin{equation}\label{p1}
 0<p<p_1:={\textstyle \frac{1}{7}} \left({\textstyle\frac{11}{2}} \log \left({\textstyle\frac{14}{11}}\right)+{\textstyle\frac{3}{2}} \log
   \left({\textstyle\frac{14}{3}}\right)\right)
   \approx 0.51958
 \;,
\end{equation}
see Figure \ref{FIGSS8}.
For the value $p=p_1$ the initial state $\rho(p_1)$ is mapped under $\mathfrak{I}({\mathcal N})$ onto $\rho(1-p_1)$
which has the same entropy.
Recall that within the family $\rho(p),\;0\le p\le p_1$ only states with $0<p<1/2$ have a positive temperature
and hence for these entropy decrease is guaranteed.

Although the instrument $\mathfrak{I}$ does not satisfy the conditions of Theorem \ref{T1} or Corollary \ref{Cor1}
its entropy decrease satisfies the same bound given in (\ref{theorem1}), see Figure \ref{FIGSS8}.

According to the considerations of Section \ref{sec:DR} and Proposition \ref{prop0} 
it is clear that a possible decrease of entropy will be
compensated by an equal or larger increase of entropy of the auxiliary system, i.~e., of the ``heat bath".
Nevertheless, it will be instructive to check this result for the considered example, see Figure \ref{FIGSS6}.
We have plotted the initial total entropy $S_0+S_i$  (red curve) and the final entropy $S_1+S_2$ (blue curve)
calculated for the one-parameter family $\rho(p)$ according to (\ref{rhop0}) such that $S_0=S(\rho(p))$.
Obviously, $S_0+S_i< S_1+S_2$ for $0<p\le 1$ in accordance with the second law.
$S_i=\log 14$ denotes the initial
entropy of the ``heat bath" due to $14$-fold degeneracy of the ground state of $H_6$.
At $p=1$ the final entropy $S_1+S_2$ assumes the value
\begin{equation}\label{Sf}
  S_f={\textstyle\frac{4}{7}} \log \left({\textstyle\frac{7}{4}}\right)+{\textstyle\frac{3}{7}} \log
   \left({\textstyle\frac{7}{3}}\right)+{\textstyle\frac{5 \log (14)}{14}}+{\textstyle\frac{4}{7}}\ \log
   \left({\textstyle\frac{63}{4}}\right)+{\textstyle\frac{\log (126)}{14}}
 \approx 3.54621
\;,
\end{equation}
see Figure \ref{FIGSS6}.

\section{Summary and Outlook}
\label{sec:SU}

In this paper we have elaborated a recent proposal \cite{S20} to describe the ``intervention of intelligent beings" in quantum systems in terms of ``conditional action". This is a genuinely physical concept. Mathematically, the notion of general ``instruments", originally intended to explain state changes due to measurements, is already broad enough to include conditional action.

A fundamental assumption here is that it is not necessary to describe the inner life of ``intelligent beings" in more detail; it is sufficient to analyze the workings of apparatuses built to realize measurements and conditional actions.
Ideally, such an analysis includes the original measurement on the system $\Sigma$, 
the storage of the measurement result in a classical memory, and the subsequent unitary time evolution of $\Sigma$ conditioned by the memory contents.
But the construction of such a complete model of a conditional action would, in my opinion, require a solution of the quantum measurement problem and hence is impossible at present.

We must therefore confine ourselves to considering physical realizations of conditional actions restricted to so-called ``measurement dilations''. These are well-known mathematical constructions \cite{BLPY16} that reduce general instruments acting on $\Sigma$ to special L\"uders instruments acting on a larger system $\Sigma + E$. These tools also open the way to understanding the (possible) entropy decrease in $\Sigma$ due to the conditional action as an entropy flow from $\Sigma$ to the auxiliary system $E$, in the same sense as the (possible) entropy decrease due to a general measurement can be explained.

The latter explanation can also be related to existing approaches to resolving the apparent contradiction of said entropy decrease with a tentative second law of quantum thermodynamics. Among such approaches are the Szilard principle, the Landauer-Bennett principle, and the recent OLR approach.
Due to the Szilard principle the entropy decrease in $\Sigma$ is, at least, compensated by the entropy production associated with the measurement
of the system's state. This principle has been confirmed by the present conditional action approach in the special case where the auxiliary system $E$
can be conceived as a memory device, see Theorem \ref{T1}, but not in general. Also a partial compatibility to the OLR approach has been shown,
in so far as, in special cases,  the entropy decrease in $\Sigma$ is bounded by the loss of mutual information due to conditional action, see Proposition \ref{prop2}.
Similarly, the approach based on the entropy costs of memory erasure (Landauer-Bennett principle) is compatible with our approach,
but cannot be viewed as the ultimate solution of the apparent paradox.

We have analyzed the imperfect erasure of a qubit by means of a physical model. This model descries the cooling of a single spin by coupling it to
a ``cold bath" consisting of six other spins such that the total time evolution can be analytically calculated. This model thus represents
a more or less realistic measurement dilatation of imperfect erasure conceived as a conditional action, and as such motivates the slight generalization of this concept compared to \cite{S20}. At the same time this example reveals some problems of the mentioned principles based on acquisition or deletion
of information, since imperfect erasure of a qubit can be achieved without any measurement at all. This is even more plausible if one considers the physical interpretation of the erasure as a cooling of a single spin. The OLR approach appears to avoid this problem because it relies on an information concept that is independent of possible measurements, see \cite{A08} for a corresponding treatment of imperfect memory erasure.

For future investigations it seems to be a desirable goal to extend the conditional action approach to the field of classical physics. First steps toward this goal restricted to discrete state spaces have been made in \cite{S20}. The role of measurement is different in classical theories because, unlike in quantum theory, there are always idealized measurements that do not change the state of the system. However, there exist non-trivial instruments describing conditional action even in classical theories, and it should be possible to realize them by extending the system analogously to the quantum case.

\appendix

\section{Construction of the standard measurement dilation for a general instrument}
\label{sec:E}

Let an instrument of the form (\ref{OI3}) be given, i.~e.,
\begin{equation}\label{E1}
 \mathfrak{J}(n)(\rho)=\sum_{i\in {\mathcal I}_n} A_{ni}\,\rho A_{ni}^\ast,\quad n\in{\mathcal N}
 \;,
\end{equation}
such that the corresponding observable $F=\left(F_n\right)_{n\in{\mathcal N}}$ is given by
\begin{equation}\label{Eobs}
  F_n= \sum_{i\in {\mathcal I}_n}  A_{ni}^\ast A_{ni},\quad \mbox{for all } n\in{\mathcal N}
  \;,
\end{equation}
satisfying $\sum_{n\in{\mathcal N}}F_n={\mathbbm 1}_{\mathcal H}$.

Following \cite{NC00} we want to explicitly construct a measurement dilation of $\mathfrak{J}$ of the form (\ref{inst2}),
see also the analogous construction for a Maxwell instrument in \cite{S20}.

To this end we define ${\mathcal N}':= \{(n,i)\left| \right. n\in{\mathcal N} \mbox{ and  } i\in {\mathcal I}_n \} $
and choose ${\mathcal K}={\mathbbm C}^{{\mathcal N}'}$ and an orthonormal basis $\left(|n i\rangle\right)_{n\in{\mathcal N,\, i\in {\mathcal I}_n}}$
in ${\mathcal K}$.
Let $\phi\in{\mathcal K}$ be one of these basis vectors, say,
$\phi=|11\rangle$.
Further, let $\left(Q_n\right)_{ n\in{\mathcal N}}$ denote the complete family of projectors in the Hilbert space ${\mathcal K}$ defined by
\begin{equation}\label{E1a}
 Q_n = \sum_{i\in {\mathcal I}_n} |ni\rangle \langle ni|\;,\quad
 \mbox{for all }  n\in{\mathcal N}
 \;.
\end{equation}

Moreover, let $\check{Q}_{ni}$ be the subspace of ${\mathcal H}\otimes {\mathcal K}$ formed by vectors of the form
$\psi\otimes |n i\rangle$ for all $ \psi\in {\mathcal H}$ and fixed $n\in{\mathcal N},\,i\in {\mathcal I}_n$ and
define a linear map $V_{11}:\check{Q}_{11} \rightarrow {\mathcal H}\otimes{\mathcal K}$ by
\begin{equation}\label{E2}
  V_{11}\left| \psi  11\right\rangle:= V_{11}\left( \psi \otimes |11\rangle\right)
  :=\sum_{n\in{\mathcal N},i\in{\mathcal I}_n}  A_{ni}\psi  \otimes |ni\rangle
  \;,
\end{equation}
for all $\psi\in {\mathcal H}$.

\begin{lemma}\label{L1}
 The map $V_{11}:\check{Q}_{11} \rightarrow {\mathcal H}\otimes{\mathcal K}$ is a partial iso\-metry, i.~e.,
 satisfies $V_{11}^\ast\,V_{11}={\mathbbm 1}_{\check{Q}_{11}}$.
\end{lemma}

Proof: Let $\varphi,\,\psi$  be two arbitrary vectors of ${\mathcal H}$ and consider the scalar products
\begin{eqnarray}
\left\langle \varphi 11\right| V_{11}^\ast V_{11} \left| \psi 11\right\rangle
&\stackrel{(\ref{E2})}{=}&\sum_{nimj}\left\langle \varphi\right|  A_{ni}^\ast\,A_{mj}\left| \psi \right\rangle
\underbrace{\left\langle ni\right| \left. mj\right\rangle}_{\delta_{nm}\delta_{ij}}\\
&=&\sum_{ni}\left\langle \varphi\right|  A_{ni}^\ast\,A_{ni}\left| \psi \right\rangle\\
&\stackrel{(\ref{Eobs})}{=}& \left\langle \varphi\right| \underbrace{\sum_n F_n}_{{\mathbbm 1}_{\mathcal H}}\left| \psi \right\rangle
=\left\langle \varphi 11 \right| \left. \psi 11\right\rangle
\;,
\end{eqnarray}
which completes the proof of Lemma \ref{L1}. \hfill$\Box$\\

Next we extend the partial isometry $V_{11}$ to a unitary operator
$V:{\mathcal H}\otimes{\mathcal K}\rightarrow {\mathcal H}\otimes{\mathcal K}$.
This completes the definition of the quantities ${\mathcal K},\phi,V,Q$ required for the measurement dilation.
It remains to show that $\mathfrak{J}={\mathfrak D}_{{\mathcal K},\phi,V,Q}$. To this end we
introduce an orthonormal basis $\left(\left| \ell \right\rangle\right)_{\ell=1,\ldots, d} $
in ${\mathcal H}$ and write
\begin{equation}\label{E4}
 \rho=\sum_{k \ell}|k \rangle \langle k | \rho |\ell\rangle \langle \ell|
 \;.
\end{equation}
Hence
\begin{equation}\label{E5}
 \rho\otimes P_\phi=\sum_{k \ell}|k 11\rangle \langle k | \rho | \ell\rangle \langle \ell 11|
 \;,
\end{equation}
and, further,
\begin{eqnarray}
\label{E6a}
 V\left(\rho\otimes P_\phi\right)V^\ast
 &\stackrel{(\ref{E5})}{=}& \sum_{k \ell}V\,|k 11\rangle \langle k | \rho | \ell\rangle \langle \ell 11|\,V^\ast \\
  \nonumber
   &\stackrel{(\ref{E2})}{=}&\sum_{k \ell n i m j}  A_{ni}|k\rangle \langle k| \rho | \ell\rangle \langle \ell | A_{mj}^\ast
   \otimes |n i\rangle \langle m j|\\
   \label{E6b}
   &\stackrel{(\ref{E4})}{=}& \sum_{n i m j}  A_{ni}\, \rho\,  A_{mj}^\ast    \otimes |n i\rangle \langle m j|
  \;.
\end{eqnarray}
Using
\begin{equation}\label{E7}
Q_r|n i\rangle\langle m j|Q_r=\delta_{r n}\,\delta_{r m}\,|r i\rangle\langle r j|
\;,
\end{equation}
for all $r\in{\mathcal N}$, (\ref{E6b}) implies
\begin{eqnarray}
 \label{E8}
\left({\mathbbm 1}\otimes Q_r \right)  V\left(\rho\otimes P_\phi\right)V^\ast \left({\mathbbm 1}\otimes Q_r \right)
  &=& \sum_{ij}\left(
    A_{ri}\, \rho\,  A_{rj}^\ast
  \right)
  \otimes |r i\rangle\langle r j|
  \;,
\end{eqnarray}
and
\begin{eqnarray}
\label{E9a}
{\mathcal D}_{{\mathcal K},\phi,V,Q}(r)(\rho)
  &=&\mbox{Tr}_{\mathcal K}
  \left(\left({\mathbbm 1}\otimes Q_r \right)  V\left(\rho\otimes P_\phi\right)V^\ast \left({\mathbbm 1}\otimes Q_r \right)\right)\\
  &\stackrel{(\ref{E8})}{=}&
 \sum_{ij}\left(
    A_{ri}\, \rho\,  A_{rj}^\ast
  \right)
  \mbox{Tr}_{\mathcal K}\left(|r i\rangle\langle r j|\right)\\
   \label{E9c}
  &=&
  \sum_{i}
    A_{ri}\, \rho\,  A_{ri}^\ast
   \;,
\end{eqnarray}
since $\mbox{Tr}_{\mathcal K}\left(|r i\rangle\langle r j|\right)=\delta_{ij}$ for all $r\in{\mathcal N}$.
The latter expression equals
\begin{equation}\label{E0d}
\nonumber
  \mathfrak{J}(r)(\rho)\stackrel{(\ref{E1})}{=}\sum_{i\in{\mathcal I}_r}     A_{ri}\, \rho\,  A_{ri}^\ast
   \;,
\end{equation}
thereby proving that the above construction is a correct measurement dilation of ${\mathfrak J}$.\\

\section{Proofs}
\label{sec:PR}

\subsection{Proof of  Proposition \ref{prop1}}
\label{sec:PR1}

We define
\begin{equation}\label{rhodef}
 \rho_1:=\mathfrak{I}({\mathcal N})(\rho)=\sum_{n\in{\mathcal N}}\mbox{Tr}_{\mathcal K}
  \left( \left({\mathbbm 1} \otimes Q_n\right)\rho'  \left({\mathbbm 1} \otimes Q_n\right)\right)
  \;,
\end{equation}
where
\begin{equation}\label{rhop}
 \rho':=V\left( \rho\otimes \sigma\right) V^\ast
  \;,
\end{equation}
and will prove Proposition \ref{prop1} by showing that $\rho_1= \mbox{Tr}_{\mathcal K}\left( \rho'\right)$.
Assume some orthonormal basis $\left( \ldots,|\alpha \rangle, \ldots, |\beta \rangle,\ldots\right)$ in
${\mathcal H}$ and another orthonormal basis $\left( | m \rangle\right)_{m\in{\mathcal M}}$ in ${\mathcal K}$
that is adapted to the environment observable $Q$ in the sense that
\begin{equation}\label{Qn}
  Q_n= \sum_{m\in{\mathcal M}_n} |m\rangle \langle m |
  \;,\quad \mbox{for all } n\in   {\mathcal N},
\end{equation}
w.~r.~t.~a partition ${\mathcal M}=\biguplus_{n\in{\mathcal N}}{\mathcal M}_n$ of the index set  ${\mathcal M}$.
It follows that
\begin{equation}\label{betam}
 \left({\mathbbm 1} \otimes Q_n\right)\left|\beta m \right\rangle =|\beta \rangle \otimes
 \left\{  \begin{array}{r@{\quad : \quad} l}
                        |m\rangle & m\in{\mathcal M}_n,\\
                        0 & \mbox{else},
          \end{array}
 \right.,
\end{equation}
for all $m\in{\mathcal M}$ and any base vector $\left|\beta\right\rangle$
and analogously for $\left\langle \alpha m\right|\left({\mathbbm 1} \otimes Q_n\right) $.
Hence an arbitrary matrix element of $\rho_1$ assumes the form
\begin{eqnarray}
\label{partialtrace1}
  \left\langle \alpha \right| \rho_1 \left| \beta\right\rangle
  &\stackrel{(\ref{rhodef})}{=}&
  \sum_{n\in{\mathcal N}}  \sum_{m\in{\mathcal M}}
  \left\langle \alpha m\right|  \left({\mathbbm 1} \otimes Q_n\right) \rho'  \left({\mathbbm 1} \otimes Q_n\right)\left| \beta m \right\rangle\\
\label{partialtrace2}
&\stackrel{(\ref{betam})}{=}& \sum_{n\in{\mathcal N}}  \sum_{m\in{\mathcal M}_n}\left\langle \alpha m\right|\rho'\left| \beta m \right\rangle\\
\label{partialtrace3}
&=&  \sum_{m\in{\mathcal M}}\left\langle \alpha m\right|\rho'\left| \beta m \right\rangle\\
\label{partialtrace4}
&=& \left\langle \alpha \right| \mbox{Tr}_{\mathcal K}\left( \rho'\right)\left| \beta \right\rangle
\;,
\end{eqnarray}
thereby completing the proof of Proposition \ref{prop1}.
\hfill$\Box$\\

\subsection{Proof of  Theorem \ref{T1}}
\label{sec:PR2}

For a pure instrument, the construction of a standard dilation given in Appendix \ref{sec:E}
is simplified by omitting the indices $i,j\in {\mathcal I}_n$ and the corresponding sums. Especially we obtain
\begin{equation}\label{rho12}
 \rho_{12}:=\sum_r \left({\mathbbm 1}\otimes Q_r \right)  V\left(\rho\otimes P_\phi\right)V^\ast \left({\mathbbm 1}\otimes Q_r \right)
 \stackrel{(\ref{E8})}{=}\sum_r \left(    A_{r}\, \rho\,  A_{r}^\ast   \right)  \otimes |r\rangle\langle r|
 \;.
\end{equation}
It follows that
\begin{equation}\label{Srho12}
S(\rho)\le S(\rho_{12})
\;,
\end{equation}
since the von Neumann entropy vanishes for pure state like $P_\phi$, is additive for tensor products and invariant under unitary transformations.
Moreover, it is non-decreasing under L\"uders measurements. Further we consider the two reduced states of $\rho_{12}$:
\begin{eqnarray}
\label{rho1r}
  \rho_1 &=& \mbox{Tr}_{\mathcal K}\, \rho_{12}\stackrel{(\ref{rho12})}{=}\sum_r A_r\,\rho \,A_r^\ast,\\
\label{rho2r}
  \rho_2 &=& \mbox{Tr}_{\mathcal H}\, \rho_{12}\stackrel{(\ref{rho12})}{=}\sum_r \mbox{Tr}\left( A_r\,\rho \,A_r^\ast\right) \,|r\rangle\langle r|
 \stackrel{(\ref{eff1})}{=}\sum_r \mbox{Tr}\left( \rho \,F_r\right) \,|r\rangle\langle r|=\sum_r p_r\,|r\rangle\langle r|
  \;.
\end{eqnarray}
Due to the subadditivity of the von Neumann entropy, see \cite{NC00}, 11.3.4, we conclude
\begin{equation}\label{S12}
  S(\rho)\stackrel{(\ref{Srho12})}{\le} S(\rho_{12})\le S(\rho_1)+S(\rho_2)
  \;,
\end{equation}
see also Proposition \ref{prop0}, 
and hence
\begin{equation}\label{DS}
 \Delta S = S(\rho)-S(\rho_1) \le  S(\rho_2)=H(\rho,F)
 \;.
\end{equation}
The latter equation follows from the spectral composition of $\rho_2$ due to (\ref{rho2}) which implies
\begin{equation}\label{Srho2}
 S(\rho_2)= -\sum_r p_r\,\log p_r \stackrel{(\ref{ShannonExp})}{=} H(\rho,F)
 \;,
\end{equation}
thereby completing the proof of Theorem \ref{T1}. \hfill$\Box$\\


\subsection{Proof of  Corollary \ref{Cor1}}
\label{sec:PR3}

Turning to the proof of Corollary \ref{Cor1} we assume that the  instrument $\mathfrak{I}$ can be
written as a convex sum of pure instruments, i.~e.,
\begin{equation}\label{PR1}
 \mathfrak{I}(n)(\rho)
 =\sum_{i\in I} \lambda_i\,  \mathfrak{I}^{(i)}(n)(\rho)
 \;,
\end{equation}
for all $n\in{\mathcal N}$ and $\rho\in B_1^+({\mathcal H})$, such that
\begin{equation}\label{lamdai}
 \lambda_i >0 \quad \mbox{for all } i\in I \quad\mbox{and}\quad \sum_{i\in I}\lambda_i=1
 \;.
\end{equation}

We will apply Theorem \ref{T1} for each pure instrument $ \mathfrak{I}^{(i)}$ and obtain
\begin{equation}\label{PR5}
 \Delta S^{(i)} = S(\rho)-S\left(\rho_1^{(i)}\right)\le H\left(p^{(i)}\right)
 \;,
\end{equation}
using some self-explaining notation. In particular, the Shannon entropy $H\left(p^{(i)}\right)$
is calculated for the probability distribution
\begin{equation}\label{PR6}
 p_n^{(i)}=\mbox{Tr}\left( \mathfrak{I}^{(i)}({n})(\rho)\right)
  \;,
\end{equation}
satisfying
\begin{equation}\label{PR7}
 \sum_n  p_n^{(i)}=1
 \;.
\end{equation}
Moreover,
\begin{equation}\label{PR8}
 \sum_i \lambda_i\,p_n^{(i)}
 \stackrel{(\ref{PR1},\ref{PR6})}{=} \mbox{Tr}\left(\mathfrak{I}(n)(\rho)\right)=p_n
 \;,
\end{equation}
for all $n\in{\mathcal N}$. Due to concavity of the Shannon entropy, see \cite{NC00}, Ex.~11.21, (\ref{PR8}) implies
\begin{equation}\label{PR9}
 H(\rho,F)=H(p)\ge \sum_i \lambda_i\,H\left(p^{(i)}\right)
 \;.
\end{equation}
Similarly, concavity of the von Neumann entropy, see \cite{NC00}, 11.3.5., yields
\begin{equation}\label{PR10}
  S(\rho_1)=S\left(\mathfrak{I}({\mathcal N})(\rho)\right)\stackrel{(\ref{PR1})}{=}
  S\left(\sum_i \lambda_i \mathfrak{I}^{(i)}({\mathcal N})(\rho)\right)
  \ge \sum_i \lambda_i S\left( \mathfrak{I}^{(i)}({\mathcal N})(\rho)\right)=
  \sum_i \lambda_i S\left( \rho_1^{(i)}\right)
  \;.
\end{equation}
Finally,
\begin{equation}\label{PR11}
 \Delta S =S(\rho) -S\left( \rho_1\right)\stackrel{(\ref{PR10})}{\le}
 S(\rho) -\sum_i \lambda_i\,  S\left( \rho_1^{(i)}\right)
{=}\sum_i \lambda_i\,\left(  S(\rho) - S\left( \rho_1^{(i)}\right) \right)
\stackrel{(\ref{PR5})}{\le}\sum_i \lambda_i\,H\left(p^{(i)} \right)\stackrel{(\ref{PR9})}{\le}H(p)
\;,
\end{equation}
thereby completing the proof of Corollary \ref{Cor1}. \hfill$\Box$\\

\subsection{Proof of  Proposition \ref{prop2}}
\label{sec:PR4}
By setting $A_n= U_n\,P_n$, resp. $A_n= P_n$, for all $n\in{\mathcal N}$, we obtain from (\ref{rho12}):
\begin{equation}\label{rhoP1}
 \rho_{12}=\sum_{n} \left(U_n P_n \rho P_n U_n^\ast \right)\otimes |n\rangle \langle n|=: \sum_n p_n\,\rho_n \otimes |n\rangle \langle n|
 \;,
\end{equation}
where $p_n:=\mbox{Tr} \left(\rho\,P_n\right)$,
and
\begin{equation}\label{rhoP2}
 \rho_{12}'=\sum_{n} \left( P_n \rho P_n  \right)\otimes |n\rangle \langle n|
 \;.
\end{equation}
 Moreover, by means of (\ref{rho2}),
\begin{equation}\label{rho2p}
 \rho_2 = \rho_2' = \sum_n p_n |n\rangle \langle n|
 \;,
\end{equation}
and hence
\begin{equation}\label{S2p}
 S_2=S_2' = H(p)=H(\rho, P)
 \;.
\end{equation}
Since the $\rho_n \otimes |n\rangle \langle n|$  in (\ref{rhoP1}) as well as the
$\left( P_n \rho P_n  \right)\otimes |n\rangle \langle n|$ in (\ref{rhoP2})
have orthogonal support, theorem 11.8 (4) of  \cite{NC00} can be applied and yields:
\begin{eqnarray}
\label{S12a}
 S_{12} &=& S\left(\rho_{12}\right)=\sum_n p_n S\left( \rho_n\otimes |n\rangle \langle n|\right)+H(p) \\
 \label{S12b}
   &=& \sum_n p_n S\left( \frac{1}{p_n}U_n P_n \rho P_n U_n^\ast\right) +H(p)\\
   \label{S12c}
   &=&\sum_n p_n S\left( \frac{1}{p_n} P_n \rho P_n \right) +H(p)\\
   \label{S12d}
   &=& S_{12}'
   \;,
\end{eqnarray}
using the invariance of von Neumann entropy under unitary transformations in (\ref{S12c}).
Finally,
\begin{eqnarray}
\label{DIa}
  \Delta {\mathcal I} &\stackrel{(\ref{OLRa}-\ref{OLRc})}{=}& \left(S_1'+S_2'-S_{12}' \right)- \left(S_1+S_2-S_{12} \right)\\
  \label{DI}
   &\stackrel{(\ref{S12d})}{=}& S_1'+S_2'-S_1-S_2 \stackrel{(\ref{S2p})}{=}S_1'-S_1  \\
   \label{DIc}
   &\ge& S_0-S_1 =\Delta S
   \;,
\end{eqnarray}
where we have used $S_1'\ge S_0$ in (\ref{DIc}) since a total L\"uders operation never decreases entropy.
This completes the proof of Proposition \ref{prop2}. \hfill$\Box$\\

\begin{acknowledgments}
I thank all members of the DFG Research Unit FOR 2692
as well as Thomas Br\"ocker for stimulating and
insightful discussions.
\end{acknowledgments}


\end{document}